\documentclass[aps,prl,reprint,showpacs,byrevtex]{revtex4-2}
\let\oldhat\hat
\renewcommand{\hat}[1]{\oldhat{\mathbf{#1}}}
\usepackage{titlesec}
\usepackage{array}

\usepackage{times,mathptm}
\usepackage[dvips]{graphicx,color}

\begin{document}
\title{Emerging Two-Dimensional Magnetism in Nonmagnetic Electrides Hf$_2$X (X = S, Se, Te)}
\author{Shuyuan Liu$^1$, Chongze Wang$^1$, Hyunsoo Jeon$^1$, Jia Yu$^2$, and Jun-Hyung Cho$^{1*}$}
\affiliation{$^1$Department of Physics and Research Institute for Natural Science, Hanyang University, 222 Wangsimni-ro, Seongdong-Ku, Seoul 04763, Republic of Korea \\
$^2$Key Laboratory for Special Functional Materials of the Ministry of Education, Henan University, Kaifeng 475004, People's Republic of China}
\date{\today}

\begin{abstract}
Recent experimental discoveries of two-dimensional (2D) magnets have triggered intense research activities to search for atomically thin magnetic systems. Using first-principles calculations, we predict the emergence of 2D magnetism in the monolayers (MLs), few layers, and surfaces of nonmagnetic layered electrides Hf$_2$X (X = S, Se, Te) consisting of three-atom-thick Hf$-$X$-$Hf stacks. It is revealed that each bulk Hf$_2$X hosts a novel quantum state of Dirac nodal lines with a high density of states arising from Hf-5$d$ cationic and interlayer anionic electrons around $-$0.9 eV below the Fermi level $E_F$. However, for the MLs, few layers, and surfaces of Hf$_2$X, such hybridized states are shifted toward $E_F$ to generate van Hove singularities, leading to a Stoner instability. The resulting surface ferromagnetism gives rise to strongly spin-polarized topological surface states at Hf$_2$X(001), demonstrating that anionic electrons, 2D magnetism, and band topology are entangled with each other. Our findings will open new perspectives for the discovery of 2D magnets via exploiting surface effects in nonmagnetic layered electrides.
\end{abstract}
\maketitle


A wide variety of 2D materials have been explored to exhibit unique physical and chemical properties that are strikingly different from those of their 3D parent compounds~\cite{2Dreview1,2Dreview2}. For examples, in contrast to graphite that is a semimetal with an overlap between the conduction and valence bands~\cite{graphite}, graphene is a zero-gap semiconductor~\cite{graphene}; some transition metal dichalcogenides (TMDCs) exhibit a change in the electronic structure from indirect bandgap in their bulk form to direct bandgap in monolayers (MLs)~\cite{TMDC}. By exploiting such dimensionality-driven novel electronic and optical properties, 2D materials including graphene, TMDCs, and the families of monoelemental (e.g., black phosphorus~\cite{phosphorus}, arsenene~\cite{arsenene}, antimonene~\cite{antimonene}, and bismuthine~\cite{bismuthine}) and ternary (e.g., Bi$_2$O$_2$Se~\cite{Bi2O2Se}, BiOX~\cite{BiOX}, and CrOX (X = Cl, Br, I)~\cite{CrOX}) compounds have been illustrated to provide many exciting new opportunities for diverse technical applications at the atomic and nanometer scales~\cite{2Dapp1,2Dapp2,2Dapp3}.

Nevertheless, 2D magnetism has been a long-standing elusive issue~\cite{2Dmagreview}. According to the Mermin-Wagner theorem~\cite{Mermin-WagnerPRL1966}, 2D magnetic systems described by the isotropic Heisenberg model cannot have a long-range magnetic ordering at any finite temperature due to thermal fluctuations. However, magnetocrystalline anisotropy enables the suppression of such thermal fluctuations, thereby allowing the stabilization of 2D magnetism~\cite{magnetocrystalline}. Despite the early availability of magnetic van der Waals (vdW)-layered crystals, the discovery of 2D magnetism in their ML or few-layer form has only recently been made experimentally~\cite{CrI3Nature,Cr2Ge2Te6Nature,FePS3-2Dmat,VSe2Nature,VSe2ACSNano}. For the vdW-layered ferromagnets CrI$_3$~\cite{CrI3Nature} and Cr$_2$Ge$_2$Te$_6$~\cite{Cr2Ge2Te6Nature}, ferromagnetic (FM) order was observed to be maintained down to the ML and bilayer limits at low temperatures, respectively. Meanwhile, bulk FePS$_3$ having an antiferromagnetic (AFM) order was experimentally observed to preserve its AFM property up to ML and few layer~\cite{FePS3-2Dmat}. Interestingly, for VSe$_2$, bulk does not have spontaneous magnetization, but ML VSe$_2$ exhibits a FM order~\cite{VSe2Nature,VSe2ACSNano}. These experimental evidences of 2D magnetism in vdW-layered materials have stimulated many searches for a variety of 2D magnetic candidate materials~\cite{2DMagnet2021}.

As the unconventional class of ionic compounds, 2D layered electrides A$_2$X consisting of a three-atom-thick building block of A$-$X$-$A stacks [see Fig. 1(a)] have recently been discovered to offer promising opportunities for both fundamental research and technological applications~\cite{catalysis-Nat.Chem2012,Ca2N-Nature2013,Electride-PRX2014,Seho,Topological-Electride-PRX2018,LiangLiu,Gd2C-PRL2020,Gd2O-PRB2022}. Depending on the cationic constituent atoms of A$_2$X, anionic excess electrons confined in the interstitial spaces between positively charged A$-$X$-$A stacks are distributed in different degrees of localization, which in turn lead to nonmagnetic (NM) or magnetic electrides~\cite{shuyuan-jpcc}. So far, NM Ca$_2$N~\cite{Ca2N-Nature2013}, Hf$_2$S~\cite{Hf2S-Sci. Adv.2020}, Sr$_3$CrN$_3$~\cite{SrCrN3}, Sr$_8$P$_5$~\cite{SrP}, and Sr$_5$P$_3$~\cite{SrP}, paramagnetic Y$_2$C~\cite{Electride-Y2C2014,Y2CFM-JACS2017,2018Y2C}, and FM Gd$_2$C~\cite{Gd2C-Nat.Commun.2020} and YCl~\cite{YCl-exp-2021} have been synthesized experimentally. Here, we theoretically predict the emergence of 2D magnetism in the MLs, few layers, and surfaces of the bulk nonmagnetic layered electrides Hf$_2$X (X = S, Se, Te). Unlike the above-mentioned 2D vdW-layered magnets, the MLs and few layers of Hf$_2$X show strong surface effects that significantly modify their band structures and charge distributions at the outermost Hf layers, thereby inducing a Stoner instability, as will be demonstrated later.

In this Letter, we focus on a recently synthesized~\cite{Hf2S-Sci. Adv.2020} electride Hf$_2$S to explore 2D magnetism in its ML, few layer, and surface using first-principles density-functional theory (DFT) calculations. We find dramatic changes in the electronic structure of ML and few layer: i.e., bulk Hf$_2$S has a large density of states (DOS) around $-$0.9 eV below the Fermi level $E_F$, arising from hybridized Hf-5$d$ cationic and interlayer anionic states, while ML and few layer Hf$_2$S exhibit a shift of such a van Hove singularity (vHs) toward $E_F$. As a result, the outermost Hf layers in the top and bottom surfaces of ML and few layer have an in-plane FM order with opposite spin polarizations. Moreover, we find that the Hf$_2$S(001) surface hosts nontrivial topological surface states associated with the bulk Dirac nodal lines, which are strongly spin-polarized to exhibit a surface ferromagnetism. The present findings can also be applicable to other isoelectronic NM electrides Hf$_2$Se and Hf$_2$Te, where emerging 2D magnetism is enhanced compared to Hf$_2$S. Therefore, we propose a new family of layered electrides Hf$_2$X (X = S, Se, Te) showing an intriguing surface-driven transformation from bulk nonmagnetic to 2D magnetic order.

We begin by examining the ground state of bulk Hf$_2$S using DFT calculations~\cite{method}. Our spin-polarized calculations for bulk Hf$_2$S show that any initial FM or AFM configuration converges to the NM state. Therefore, bulk Hf$_2$S has a NM ground state, consistent with a combined experimental and DFT study of Kang $et$ $al$.~\cite{Hf2S-Sci. Adv.2020}. Figures 1(a) and 1(b) show the optimized ground structure with the lattice parameters $a_1$ = $a_2$ = 3.375 {\AA} and $a_3$ = 11.764 {\AA}, where S atoms locating in a triangular lattice are surrounded by six Hf atoms in an trigonal prismatic geometry with the space group P6$_3$/$mmc$ (No. 194). The calculated band structure of bulk Hf$_2$S shows that Hf-5$d$ cationic and interstitial anionic states are strongly hybridized around $-$0.9 eV below $E_F$~\cite{PED}, giving rise to a large peak in their partial density of states (PDOS) [see Fig. 1(c)]~\cite{PDOS}. Here, the anionic electrons localized at the positions marked as $X_1$ and $X_2$ in the interlayer space are well represented by the electron localization function (ELF) [see Fig. 1(d)]. These local maxima positions of ELF agree well with those of a previous DFT calculation~\cite{Hf2S-Sci. Adv.2020}. Interestingly, for the ML and few layer of Hf$_2$S, such hybridized Hf-5$d$ cationic and interstitial anionic states are shifted toward $E_F$, thereby inducing a magnetic instability, as discussed below.

\begin{figure}[h!t]
\includegraphics[width=8.5cm]{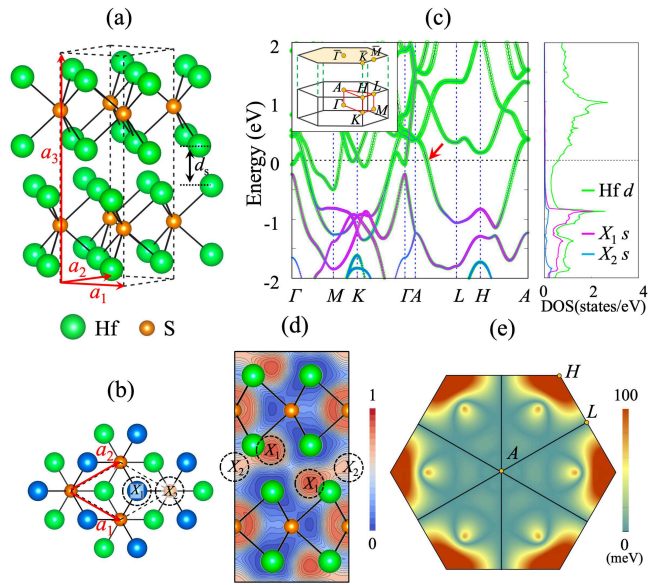}
\caption{(a) Optimized structure of bulk Hf$_2$S and (b) its top view. The primitive unit cell contains two Hf$-$S$-$Hf stacks with an alternative AB stacking sequence. In (b), the blue and green circles represent Hf atoms in two different stacks, while the dashed circles represent $X_1$ and $X_2$ anions locating at hollow sites. The calculated band structure of bulk Hf$_2$S is displayed in (c) together with the PDOS, where the projected bands onto Hf-5$d$ and $X_1$-, $X_2$-$s$-like orbitals are represented by circles whose radii are proportional to the weights of the corresponding orbitals. The inset of (c) shows the BZ of the primitive unit cell. In (d), the calculated ELF of Hf$_2$S is drawn on the (110) plane with a contour spacing of 0.05, where the MT radius of $X_1$ and $X_2$ anions is chosen as 1 {\AA}. In (e), the SOC gap of 2D nodal surfaces is displayed throughout the $k_z$=${\pi}$/$c$ plane, where the black lines represent DNLs.}
\label{figure:1}
\end{figure}

It is noticeable that there is a fourfold degenerate band crossing $E_F$ along the high-symmetry $A-L-H-A$ paths, indicated by the arrow in Fig. 1(c). Using the tight-binding Hamiltonian with a basis of maximally localized Wannier functions~\cite{wannnier90,wanniertools}, we find the existence of 2D nodal surface crossing $E_F$ throughout the $k_z$=${\pi}$/$c$ plane on the boundary of Brillouin zone (BZ)~\cite{wannier-DFT}. This nodal surface formed by a touching of two doubly-degenerate bands is respected by the nonsymmorphic crystal symmetry $S_{2z}$, equivalent to the combination of twofold rotation symmetry $C_{2z}$ about the $z$ axis and a half translation along the $z$ direction (see symmetry analysis in the Supplemental Material~\cite{SM}). The inclusion of SOC lifts the fourfold degeneracy of nodal surface except along the high-symmetry paths $k_x$=0 and $k_x$=${\pm}\sqrt{3}k_y$, preserving 1D nodal lines [see Fig. 1(e)]. These Dirac nodal lines (DNLs) showing $C_{3z}$ rotation symmetry are protected by additional mirror symmetries~\cite{note-1}. Thus, bulk Hf$_2$S is characterized as a topological semimetal having DNLs crossing $E_F$.

To reveal the effect of reduced dimensionality on electronic structure, we first consider the NM phase of ML Hf$_2$S. We find that the lattice constants become $a_1$ = $a_2$ = 3.230 {\AA}, slightly smaller than those (3.375 {\AA}) of bulk Hf$_2$S. As shown in Fig. 2(a), ML Hf$_2$S has a large peak in the PDOS of Hf-5$d$ cationic and interstitial anionic states around $E_F$. We increase the interlayer spacing $d_s$ [see Fig. 1(a)] in bulk Hf$_2$S to examine the change of band structure. As $d_s$ increases, the hybridized Hf-5$d$ cationic and interstitial anionic states locating around $-$0.9 eV are shifted toward $E_F$ (see Fig. S4 in the Supplemental Material~\cite{SM}), converging to the band structure of ML Hf$_2$S. We also find that the distribution of interstitial anionic electrons changes with respect to $d_s$: i.e., for bulk Hf$_2$S, the number of electrons $n_{X_1}$ ($n_{X_2}$) within the muffin-tin (MT) sphere of the $X_1$ ($X_2$) anion is 1.120 (0.405) electrons per ML [see Fig. 1(d)], which decreases (increases) to 0.904 (0.510) electrons in ML Hf$_2$S. Such rearrangements of $n_{X_1}$ and $n_{X_2}$ in ML Hf$_2$S [see Fig. 2(b)] together with the shift of hybridized Hf-5$d$ cationic and interstitial anionic states reflect strong surface effects, as will be demonstrated at the Hf$_2$S(001) surface. As shown in Fig. S5, the partial charge distributions with respect to the energy ranges exhibit a strong variation of anionic electrons between bulk and ML Hf$_2$S, due to breaking bonds at surfaces~\cite{note-surface}. In these senses, surface formation significantly changes the distribution of interstitial anionic electrons as well as their electronic band structure, both of which are the peculiar features of Hf$_2$X. These dramatic changes of interstitial anionic states between bulk and surface induce the emergence of 2D ferromagnetism in Hf$_2$X, as discussed below.

\begin{figure}[h!t]
\includegraphics[width=8.5cm]{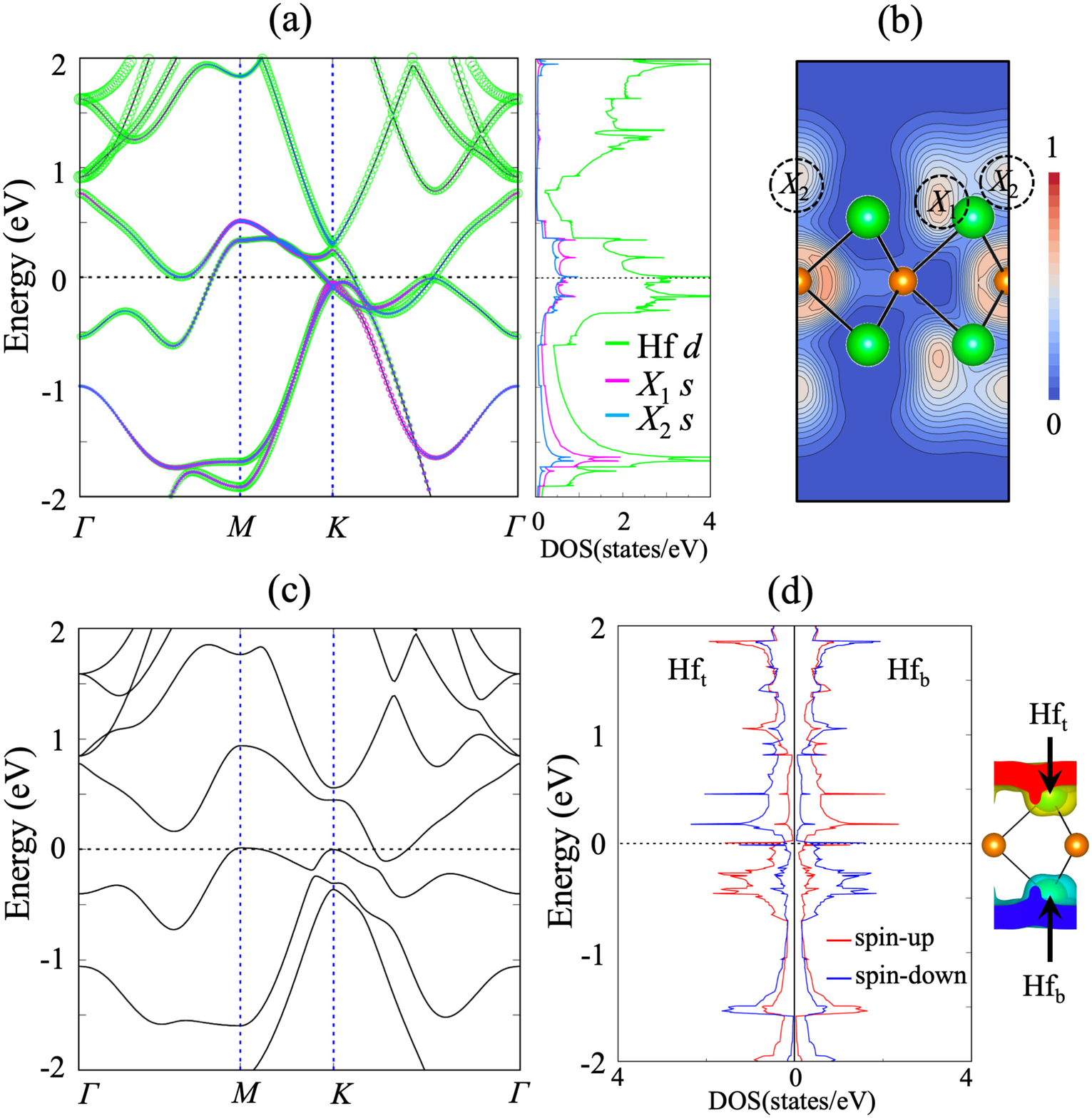}
\caption{(a) Calculated band structure with PDOS and (b) ELF (with a contour spacing of 0.05) of the NM phase of ML Hf$_2$S. The band structure and spin-polarized LDOS (projected onto Hf$_{\rm t}$ and Hf$_{\rm b}$) of the AFM phase of ML Hf$_2$S are displayed in (c) and (d), respectively. The AFM spin density is also given in (d) with an isosurface of 0.01 $e$/{\AA}$^3$.}
\label{figure:2}
\end{figure}

Since the band structure of the NM phase of ML Hf$_2$S has a vHs at $E_F$ [see Fig. 2(a)], the Stoner criterion may be fulfilled to lead to a FM instability. Indeed, we find that the FM phase is favored over the NM phase by 1.55 meV per Hf atom. However, the AFM coupling of the ferromagnetically ordered magnetic moments on the outermost Hf layers is further stabilized compared to the FM phase by 11.41 meV per Hf atom. In such an AFM ground state, the calculated spin magnetic moments integrated within the MT spheres around Hf~\cite{note-radius}, $X_1$, and $X_2$ are 0.498, 0.104, and 0.105 ${\mu}_B$, respectively. Here, the AFM coupling of Hf 5$d$ spins is likely driven by superexchange interactions~\cite{superexchange1950,GKA-Goodenough,GKA-Kanamori} through the occupied S 3$p$ states (see Fig. S6 in the Supplemental Material~\cite{SM}). Consequently, this AFM ordering opens a pseudogap for the electronic states around $E_F$ [see Fig. 2(c)]. As shown in Fig. 2(d), the spin-up and spin-down local DOS (LDOS) projected onto Hf$_{\rm t}$ and Hf$_{\rm b}$ residing at the top and bottom layers exhibit the separation between occupied and unoccupied states~\cite{note-superexchange}. By including SOC, the easy axis points along the $z$ direction with a magnetic anisotropy energy of 0.76 meV per Hf atom, indicating that ML Hf$_2$S has the Ising anisotropy with a strong AFM interlayer interaction. Since this AFM ML structure is revealed to be thermodynamically stable~\cite{MD-note} [see Figs. S7(a) and S7(b)], we anticipate that it would be experimentally synthesized in the future by either mechanical exfoliation such as graphene~\cite{ex-graphene} and MoS$_2$~\cite{ex-mos2} or epitaxial growth on proper substrates such as silicene~\cite{growth-silicene}, stanene~\cite{growth-stanene}, and tellurene~\cite{growth-tellurene}.


Next, we investigate the stability of the FM and AFM phases of few-layer Hf$_2$S with increasing the number $N$ of Hf$-$S$-$Hf stacks. Figure 3(a) shows that the energy difference ${\Delta}E$ between the FM and AFM phases decreases sharply even at $N$ = 2, indicating that the top and bottom surfaces of few-layer Hf$_2$S can form an isolated FM order with their suppressed AFM coupling. Figure 3(b) shows the spin density at the (001) surface, obtained using a periodic slab of $N$ = 12 with ${\sim}$25 {\AA} of vacuum in-between adjacent slabs. We find that the spin magnetic moment exists mostly at the topmost Hf layer, while it is significantly reduced at the second and third Hf layers (see Table I). The projected LDOS demonstrates that the spin-up and -down states arising from Hf-1 atom are separated by ${\sim}$0.47 eV~\cite{note-split}, while those from Hf-2 atom exhibit a little separation. By dividing this exchange splitting of Hf-1 by the corresponding magnetic moment, we can estimate the Stoner parameter $I$, which satisfies the Stoner criterion $D(E_{F})I > 1$~\cite{stoner} [see Fig. S8(a) in the Supplemental Material~\cite{SM}]. Here, $D(E_{F})$ is the DOS at $E_F$ from the NM phase. Thus, we can say that surface ferromagnetism emerging at the Hf$_2$S(001) surface is driven by the Stoner instability due to a vHs at $E_F$ arising from hybridized Hf-5$d$ cationic and surface anionic states [see Fig. S8(b)]. In order to estimate the Curie temperature $T_{\rm c}$, we perform spin-polarized calculations for various AFM surface configurations (see Fig. S9). We find that the lowest AFM configuration is less stable than the FM one by 15.35 meV per Hf atom. Using the mean field approximation~\cite{MFA}, we estimate a $T_{\rm c}$ of ${\sim}$118 K.

\begin{figure}[h!t]
\includegraphics[width=7.5cm]{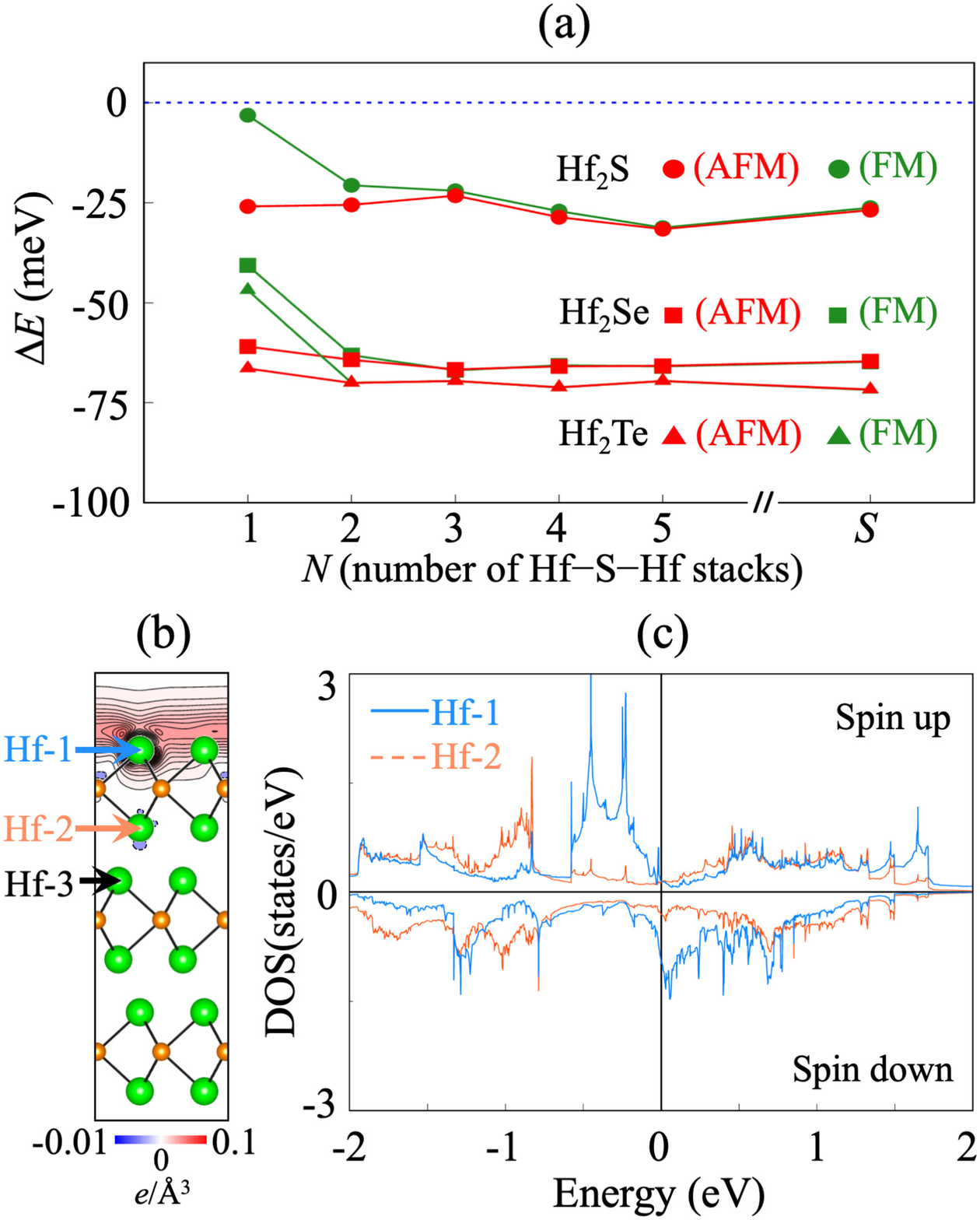}
\caption{(a) Calculated total energies of the FM and AFM phases (relative to that of the NM phase) of Hf$_2$S, Hf$_2$Se, and Hf$_2$Te as a function of $N$. Here, "$S$" represents the data of the (001) surface, obtained using a periodic slab of $N$ = 12. The spin density (with a contour spacing of 0.005 $e$/{\AA}$^3$) and the LDOS of Hf-1 and Hf-2 atoms at the Hf$_2$S(001) surface are given in (b) and (c), respectively.}
\label{figure:3}
\end{figure}

According to the bulk-boundary correspondence of topological nodal line semimetals~\cite{DNL-review}, the presence of DNLs in bulk leads to the formation of topologically protected surface states~\cite{DNL1,DNL2}. Using the Green's function method based on the tight-binding Hamiltonian with maximally localized Wannier functions~\cite{wannnier90,wanniertools}, we obtain the projected surface spectrum of the NM Hf$_2$S(001) surface [see Fig. 4(a)]. There are three drumhead surface states $SS_1$, $SS_2$, and $SS_3$ along the ${\overline{\Gamma}}-{\overline{M}}-{\overline{K}}-{\overline{\Gamma}}$ path near $E_F$, which represents the hallmark of DNLs~\cite{drumhead}. Note that $SS_1$, $SS_2$, and $SS_3$ are split by the SOC-driven gap openings (see Fig. S10 in the Supplemental Material~\cite{SM}). In Fig. 4(b), the Fermi surface of Hf$_2$S(001) at a chemical potential of $-$0.2 eV exhibits the closed Fermi arcs around the ${\overline{\Gamma}}$ and $\overline{K}$ point with helical spin textures. Such nontrivial topological surface states with a unique spin-momentum locking property~\cite{helicalspin} can be more distinguishable by subtracting the (001) projected bulk states from the surface spectrum (see Fig. S11). For the experimental measurements of these topological surface states, we propose the H-passivation of the Hf$_2$S(001) surface where surface ferromagnetism can be removed~\cite{H-structure}. Our DFT band structure of the H-passivated Hf$_2$S(001) surface reproduces the dispersion of the $SS_1$, $SS_2$, and $SS_3$ states around $E_F$ [see Figs. 4(c) and S10]. Meanwhile, the DFT band structure of a clean Hf$_2$S(001) surface shows that the surface-induced ferromagnetism gives rise to strong spin polarizations for the $SS_1$, $SS_2$, and $SS_3$ states at Hf$_2$S(001) [see Fig. 4(d)], indicating that the bulk DNLs are split into two spin-polarized nondegenerate bands due to time-reversal symmetry breaking at surface. These results reflect a strong correlation between spin degree of freedom and topological properties. It is thus likely that anionic electrons, 2D magnetism, and band topology are entangled with each other in a new class of electrides Hf$_2$X.

\begin{figure}[h!t]
\includegraphics[width=8.5cm]{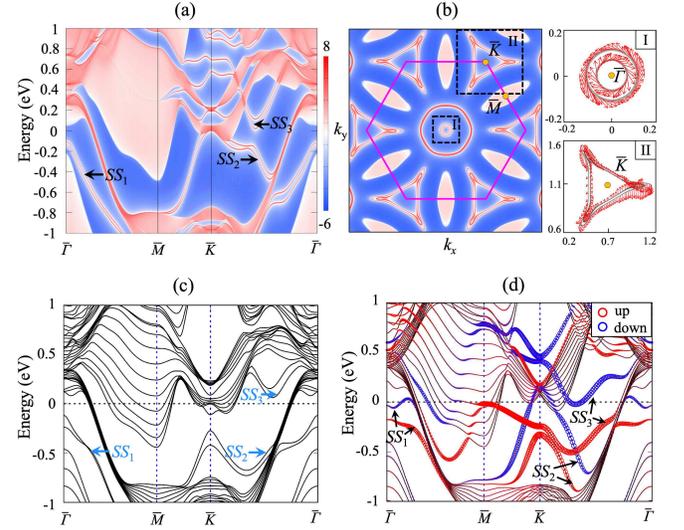}
\caption{(a) Projected surface spectrum of the NM Hf$_2$S(001) surface and (b) its isoenergy surface at $-$0.2 eV with the in-plane spin textures of the $SS_1$, $SS_2$, and $SS_3$ surface states around the ${\overline{\Gamma}}$ and ${\overline{K}}$ points. The DFT band structures of the H-passivated Hf$_2$S(001) and clean Hf$_2$S(001) surfaces are given in (c) and (d), respectively. The former (latter) bands are obtained with (without) including SOC. In (d), the radii of circles are proportional to the weights of the projection onto Hf-1 atom.}
\label{figure:4}
\end{figure}

Finally, we examine 2D magnetism in other isoelectronic NM electrides Hf$_2$Se and Hf$_2$Te. Similar to the case of Hf$_2$S, the MLs and few layers of Hf$_2$Se and Hf$_2$Te shift hybridized Hf-5$d$ cationic and interlayer anionic states toward $E_F$ (see Fig. S13 in the Supplemental Material~\cite{SM}), thereby inducing a 2D magnetism at their outermost Hf layers. As shown in Fig. 3(a), the magnetic stabilities in the MLs, few layers, and surfaces of Hf$_2$Se and Hf$_2$Te are enhanced compared to the corresponding ones of Hf$_2$S. Accordingly, the spin magnetic moments of Hf-1 atom in Hf$_2$Se and Hf$_2$Te are larger than that in Hf$_2$S (see Table I). For the Hf$_2$Se(001) and Hf$_2$Te(001) surfaces, the topological surface states associated with the bulk DNLs also exhibit the large spin splits around $E_F$ (see Fig. S14), leading to a strong surface ferromagnetism.

\begin{table}[ht]

\caption{Calculated spin magnetic moments (in unit of ${\mu}_B$ per Hf atom) of Hf-1, Hf-2, and Hf-3 at the Hf$_2$S(001), Hf$_2$Se(001), and Hf$_2$Te(001) surfaces. The values of $X_1$ and $X_2$ at surface are also given.}
\begin{ruledtabular}
\begin{tabular}{lccccc}


   & Hf-1 &  Hf-2    &  Hf-3 &$X_1$  &$X_2$ \\  \hline
Hf$_2$S(001)   &0.490 &  -0.011 &  0.001 &0.086 &  0.100 \\
Hf$_2$Se(001)  &0.511 &  -0.011 &  0.001 &0.084 &  0.092 \\
Hf$_2$Te(001) &0.536 &  -0.011 &  0.001 &0.083 &  0.083 \\
\end{tabular}
\end{ruledtabular}
\end{table}

In summary, our first-principles DFT calculations have demonstrated the importance of surface effects that invokes the emergence of 2D magnetism in the MLs, few layers, and surfaces of nonmagnetic layered electrides Hf$_2$X (X = S, Se, Te). Specifically, we revealed that nontrivial topological surface states associated with the bulk DNLs are largely spin-polarized to form a strong ferromagnetism at the Hf$_2$X(001) surfaces. Our findings provide a novel platform to investigate the intriguing interplay between electride properties, nontrivial band topology, and surface ferromagnetism, which will be promising for future spintronics technologies~\cite{app1,app2}. We found that Zr$_2$X (X = S, Se, Te) can also exhibit similar electronic structures, 2D magnetism, and topological properties [see Figs. S7(c), S7(d), S15, and Table SI], as predicted for Hf$_2$X. Thus, the present prediction of surface ferromagnetism via the strong variation of hybridized transition-metal $d$ orbitals and interstitial anionic states between bulks and surfaces of Hf$_2$X and Zr$_2$X are rather generic, thereby providing a more general physical picture to support our results.

\vspace{0.4cm}



\noindent $^{*}$ Corresponding author: chojh@hanyang.ac.kr


\begin{thebibliography}{99}
\bibitem{method} Our first-principles DFT calculations were performed using the Vienna $ab$ $initio$ simulation package with the projector-augmented wave method~\cite{vasp1,vasp2,paw}. The exchange-correlation energy was treated with the generalized-gradient approximation functional of Perdew-Burke-Ernzerhof~\cite{pbe}. The plane wave basis was employed with a kinetic energy cutoff of 550 eV, and the $k$-space integration was done with 18${\times}$18${\times}$1 and 18${\times}$18${\times}$6 meshes for ML (few layers) and bulk, respectively. All atoms were allowed to relax along the calculated forces until all the residual force components were less than 0.005 eV/{\AA}. The phonon spectrum calculations of ML Hf$_2$S were carried out by using the QUANTUM ESPRESSO package~\cite{qe}, with the 6${\times}$6${\times}$1 $q$ points.
\bibitem{PED} This unique feature of anionic electrons in bulk Hf$_2$X can be clearly seen by plotting their charge density around -0.9 eV (see Fig. S1). It is found that the integrated charge density of bulk Hf$_2$S between -0.7 and -1.4 eV describes a charge distribution of hybridized Hf-5$d$ cationic and interstitial anionic states, while that between $E_F$ and -0.7 eV represents Hf-5$d$ states.
\bibitem{PDOS} It is noted that for bulk Hf$_2$S, the PDOS for S-3$s$ and -3$p$ orbitals are located below -3 eV (see Fig. S2). Thus, S atoms hardly participate in hybridization with interstitial anionic states.
\bibitem{note-surface} It is noted that the redistribution of anionic electrons at surfaces causes the upward shift of interstitial anionic states upon dimensionality reduction. Such intriguing surface effects occurring in Hf$_2$X are absent in conventional layered electrides~\cite{Seho,LiangLiu}, the bulks of which have largely delocalized interstitial anionic states near $E_F$.
\bibitem{note-radius} The radius of the MT sphere around Hf atom is chosen as 1.6 {\AA}.
\bibitem{note-split} We estimate the exchange splitting of spin-up and -down states by calculating the average difference of their Kohn-Sham eigenvalues. Here, we considered the spin-up and -down states separately occupying 1.67 electrons, which are the number of spin-up electrons occupied below $E_F$.
\end{thebibliography}
\end{document}